\documentstyle[twocolumn,aps,psfig]{revtex}
\begin{document}
\draft
\flushbottom
\twocolumn[
\hsize\textwidth\columnwidth\hsize\csname @twocolumnfalse\endcsname

\title{ On the possibility to supercool molecular hydrogen down to superfluid
transition}
\author{Igor I. Smolyaninov }
\address{ Electrical Engineering Department \\
University of Maryland, College Park,\\
MD 20742}
\date{\today}
\maketitle
\tightenlines
\widetext
\advance\leftskip by 57pt
\advance\rightskip by 57pt

\begin{abstract}
Recent calculations by Vorobev and Malyshenko (JETP Letters, 71, 39, 2000) show 
that molecular hydrogen may stay liquid and superfluid in strong electric fields
of the order of $4\times 10^7 V/cm$. I demonstrate that strong local electric 
fields 
of similar magnitude exist beneath a two-dimensional layer of electrons 
localized
in the image potential above the surface of solid hydrogen. Even stronger local
fields exist around charged particles (ions or electrons) if surface or
bulk of a solid hydrogen crystal is statically charged. Measurements of the 
frequency
shift of the $1 \rightarrow 2$ photoresonance transition in the spectrum of
two-dimensional layer of electrons above positively or negatively charged solid 
hydrogen surface performed in the temperature range 7 - 13.8 K support the 
prediction of electric field induced surface melting. The range of surface 
charge 
density necessary to stabilize the liquid phase of molecular hydrogen at the 
temperature of superfluid transition is estimated.

%\vspace{0.5cm}

%PACS number(s): 05.70.Fh, 05.70.Ce, 05.30.Ip 
\end{abstract}

]
\narrowtext

\tightenlines

The prediction by Ginzburg and Sobyanin \cite{1} that sufficiently supercooled 
liquid
molecular hydrogen may undergo transition to a superfluid state attracts much
current attention. Recent theoretical calculations \cite{2,3} show that such a 
transition may occur
around 1.1 - 1.2 K. On the experimental side, a lot of work has been done on 
molecular hydrogen
in porous vycor glass \cite{4,5} and on the thin hydrogen films adsorbed on 
different 
substrates \cite{6}. Although some supercooling of liquid hydrogen was indeed 
observed in the experiment,
the temperatures achieved were not sufficiently low to induce superfluid 
transition.

Very recently a new possibility to produce a strongly supercooled molecular 
hydrogen
state has been suggested on the basis of thermodynamic functions calculation of 
the
molecular hydrogen in both the stable and metastable regions \cite{7}. It has 
been proposed to 
expose a two-phase system of solid and liquid hydrogen to the effect of strong 
external
electric field. According to \cite{7}, the condition of phase equilibrium in 
this case may be 
written as   

\begin{equation} 
\mu _1(p)-(\frac{\partial \epsilon _1}{\partial \rho _1})\frac{E_1^2}{8\pi} \pm 
\frac{\epsilon_1E_1^2-\epsilon _2E_2^2}{8\pi \rho _1}=\mu _2(p)-(\frac{\partial 
\epsilon _2}{\partial \rho _2})\frac{E_2^2}{8\pi } 
\end{equation}

where $\mu _1(p)$ and $\mu _2(p)$ are the chemical potentials of the solid (1) 
and liquid (2)
hydrogen phases at pressure $p$, $E$ is the electric field intensity, $\epsilon 
$ is the
dielectric constant, and $\rho $ is the density. The plus sign corresponds to 
the case when
the field is generated by constant charges, and the minus sign corresponds to 
the field
created by constant potential. The action of electric field on the two-phase 
system 
may be understood in terms of creation of "different pressures" in the phases. 
As a result, the
liquid and solid hydrogen phases which normally can not coexist at any positive 
pressure
around 1 - 2 K, may coexist in equilibrium with each other around the 
temperature of superfluid
transition at small positive pressures. Unfortunately, the electric field 
necessary for this
to occur is quite large due to the small difference in $\epsilon $ for liquid 
and solid 
hydrogen. The result obtained in \cite{7} for the case of field generated by 
constant potentials
and the field lines parallel to the phase boundary looks like

\begin{equation} 
E=(\frac{-24\pi (\mu _1(0)-\mu _2(0))\rho _1\rho _2}{(\epsilon _1^2+4\epsilon _1-2-
3\epsilon _2)(\rho _1-\rho _2)})^{1/2}=4\times 10^7 V/cm 
\end{equation}

where Clausius-Mosotti relation was used to get the value of 
$(\partial \epsilon  /\partial \rho )_T$, and the following values of dielectric 
constant and
density were accepted: $\epsilon _1=1.3$, $\epsilon _2=1.25$, $\rho _1=0.087 
g/cm^3$, and 
$\rho _2=0.078 g/cm^3$. About the same value of the electric field may be 
obtained for the
cases of field created by constant charges and/or field lines perpendicular to 
the phase
boundary. It was unclear for the authors of \cite{7} if such strong fields can 
be created in solid
hydrogen because of dielectric breakdown and other experimental difficulties.

In this Letter I am going to show that strong local electric fields of similar 
magnitude exist 
in some real physical systems created in the lab, namely, beneath a two-
dimensional layer of 
electrons localized in the image potential above the surface of solid hydrogen 
\cite{8}. Even stronger 
local fields exist around charged particles (ions or electrons) if surface or 
bulk of a solid 
hydrogen crystal is statically charged. I am going to present the data of 
measurements of the 
frequency shift of the $1 \rightarrow 2$ photoresonance transition in the 
spectrum of such
two-dimensional layer of electrons above positively or negatively charged solid 
hydrogen surface performed in the temperature range 7 - 13.8 K. These previously 
unpublished
data obtained in \cite{9} strongly support the prediction \cite{7} of electric 
field induced melting of
solid hydrogen. 
I am also going to estimate the range of surface charge density necessary to 
stabilize the liquid 
phase of molecular hydrogen at the temperature of superfluid transition.

Most of the experimental and theoretical work on two-dimensional surface image 
states above
the dielectric surfaces with $(\epsilon -1) << 1$ has been done for the case of 
liquid helium \cite{10}.
In the simplest model, the interaction potential $\phi $ for an electron near 
the surface of
such a dielectric depends only on the electrostatic image force, and on the 
external electric
field $E$, which is normal to the surface and is necessary for the electron 
confinement near
the surface:

\begin{equation} 
\phi (z)=-e^2(\epsilon -1)/(4z(\epsilon +1))+eEz=-Qe^2/z+eEz 
\end{equation}

for $z>0$, and $\phi (z)=V_0$ for $z<0$, 
where the z axis is normal to the surface, and $V_0$ is the surface potential 
barrier. If 
$V_0\rightarrow \infty$ one obtains the electron energy spectrum

\begin{equation} 
E_n=-Q^2me^4/(2\hbar ^2n^2 )+eE<z_n>+p^2/(2m)
\end{equation}

The first term in this expression gives the exact solution for $E=0$. The second 
term
is the first-order correction for a non-zero confining field, where the average 
distance
of electrons from the surface in the $n$th energy level is $<z_n>=3n^2\hbar 
^2/(2me^2Q)$.
This correction provides an extremely convenient way of fine-tuning the energy 
spectrum.
The last term corresponds to free electron's motion parallel to the surface. It 
is important 
to mention that at sufficiently
high electron density individual electrons in these surface states are no longer 
free.
When the average electrostatic potential energy per electron exceeds about 
$100k_BT$ the
electron layer undergoes transition into the two-dimensional Vigner crystal 
state \cite{10} with the 
parallel motion of individual electrons substantially restricted by other 
electrons in the 
lattice of the Vigner crystal.

Similar two-dimensional electron layers have been observed on the surface of 
liquid and solid
hydrogen. Resonance absorption of light for $1\rightarrow 2$ and $1 \rightarrow 
3$ transitions 
in the spectrum of electrons levitating above the surface of solid hydrogen has 
been
reported and the frequencies of these transitions were measured as a function of 
the confining
electric field \cite{8}. While general agreement with the spectrum (4) within 
$30\% $ has been
observed, quite a few important questions have been raised by these experiments 
which have
not yet found satisfactory answers. The most striking feature of this system is 
a very strong
dependence of the photoresonance frequencies on the hydrogen temperature (or 
vapor pressure).
The frequency of $1\rightarrow 2$ transition in zero confining field grew by $20 
\%$ on the 
cooling down from 13.6 K
to 7 K. In the absence of any other competing theory at the time of the 
measurements, this effect 
was interpreted in terms of quantum refraction \cite{11} due to the presence 
of hydrogen vapor molecules around the levitating electrons. For this 
explanation to be true,
the scattering length value for the scattering of an electron by a hydrogen 
molecule has to be
equal to $L=-1.4 \AA $, contrary to the currently accepted value of $L=+0.672 
\AA $ [12].
On the other hand, if spectral changes are completely attributed to the increase 
in the value
of $Q=(\epsilon -1)/(4(\epsilon +1))$, the frequency shift may be interpreted as 
a
gradual freezing of the supercooled liquid hydrogen film on top of the hydrogen 
crystal (with an 
increase of the dielectric constant of the thin surface layer from the liquid 
hydrogen value 
of 1.25 to the  solid hydrogen value of 1.3). This consideration raises an 
important question: 
how strong is the electric field on the surface of the solid hydrogen just below 
an electron 
in the ground surface state? Can it cause a substantial supercooling of the 
liquid molecular 
hydrogen phase on the surface?   

If we imagine for a moment that the electron is not moving parallel to the 
surface, the answer
is very easy to get. We may forget about small contribution from the electron's 
image and
write the average value of the local field as $<E_L>=<e/z_1^2>=9e/(2<z_1>^2)$, 
where we have used 
the ground state wave function for an idealized potential (3) with an infinite 
potential barrier. 
Taking into account the experimentally observed value of $<z_1>=20 \AA $ 
\cite{13}, we obtain 
$<E_L>=1.6\times 10^7 V/cm$. This field is of the same order of magnitude as the 
field that is 
(according to \cite{7}) sufficient to stabilize superfluid liquid hydrogen 
phase. 

Unfortunately, free electrons rapidly move around the surface and every region 
of solid hydrogen 
surface experiences this strong electric field for very brief periods of time. 
This situation
changes at sufficiently high electron density and low temperature when the 
electron layer 
undergoes transition into the Vigner crystal state. In this state the motion of 
each individual
electron is substantially restricted to a small area around electron's 
equilibrium position in the
Vigner crystal lattice by electrostatic interaction with other electrons in the 
lattice. Let us 
estimate the size of this area for the parameters of the electron system 
typically observed in 
the experiment.

The density of electrons on the solid hydrogen surface is determined by the 
external confining
field: $n\leq E/(4\pi e)$ \cite{8}. Density values of the order of $3\times 
10^{10} - 10^{11} cm^{-2}$ 
were routinely observed. At $n=10^{11} cm^{-2} $ an electrostatic potential 
energy per electron 
is of the order of $e^2n^{1/2}=540k_B\times 1K$. Thus, at T=1K such an electron 
system will be
in two-dimensional Vigner crystal state. Classically allowed area of electron's 
motion around
its equilibrium state in the lattice may be estimated by taking into account 
only the nearest
neighbors located at a distance $d$ from each other. For the potential energy we 
may write
approximately

\begin{equation} 
V=e^2/(d-x)+e^2/(d+x)=2e^2/d+2e^2x^2/d^3
\end{equation}

where $x$ is the displacement towards the nearest neighbor. Immediately we 
obtain an estimate
for the radius of the allowed area as $x=n^{-1/2}(k_BT/(e^2n^{1/2}))^{1/2}$. At 
T=1K and 
$n=10^{11} cm^{-2}$ this radius is approximately $10\AA $. Thus, an electron 
stays over the
same region of the hydrogen surface. We may now conclude that at the
above mentioned parameters of the electron system and at the temperatures near 
the temperature
of suspected superfluid transition a substantial portion of the solid hydrogen 
surface under
the electrons will experience strong electric fields of the order of the field 
necessary
to induce the superfluid transition. Taking into account that the maximum 
density of surface 
electrons may be substantially increased with respect to the values observed in 
\cite{8}, this
system looks extremely promising for observation of superfluid transition of the 
supercooled
molecular hydrogen. 

Another evident way to create even stronger local electric fields inside a solid 
hydrogen crystal 
is accumulation of static electric charge in the form of positive or negative 
ions and bound 
electrons on the surface or in the bulk of hydrogen crystal. Such an 
accumulation of negative
and positive static charge was a routine problem in our spectroscopic 
experiments. The detection
of static charge has been discussed in detail in \cite{8}.
The results of these experiments published so far \cite{8,13} were carefully 
selected to avoid any 
influence of static charges which were considered to be a problem. Nevertheless, 
the influence
of negative and positive static charge of the hydrogen surface on the frequency 
and the
linewidth of photoresonance $1\rightarrow 2$ transitions was carefully measured 
and reported
in \cite{9}. Because of the recent appearance of paper \cite{7} and considering 
the arguments above, 
I feel necessary to make physical community aware of these results. 

The most careful measurements of the influence of negative static surface charge 
on the
frequency and linewidth of $1\rightarrow 2$ transitions were performed at a 
fixed temperature
T=13.4K (P=40 Torr), just below the triple point of hydrogen. The negative 
surface charge was 
formed by energetic electrons which were not stopped by the surface potential 
barrier $V_0$ upon 
the deposition of free electrons on the solid hydrogen surface. Fig.1 reproduces 
the data of these 
measurements as they were presented in \cite{9}. Both the frequency shift and 
the line broadening was 
measured at a fixed wavelength of excitation light by tuning the confining field 
$E$ (that is why 
both values are expressed in the units of electric field). It is easy to convert 
these data into
units of frequency by using the spectrum (4). The frequency shift was very large 
and negative 
(corresponding to the positive change in the confining field at a fixed laser 
frequency) and much 
bigger than the line broadening. It corresponds to more than $20\% $ reduction 
of the frequency 
of $1\rightarrow 2$ transition in zero confining field. This shift did not find 
an adequate
explanation at the time of the measurements. It is consistent though with the 
picture of
supercooling of liquid hydrogen on the surface of the solid hydrogen phase in 
the presence of
static electric charges. 

\begin{figure}[tbp]
\centerline{
\psfig{figure=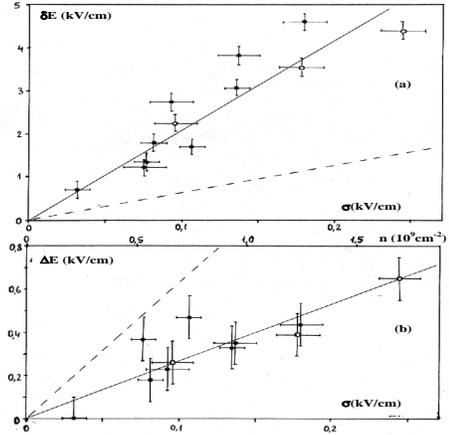,width=8.0cm,height=7.4cm,clip=}
}
\caption{ Line shift (a) and line broadening (b) of the photoresonance 
$1\rightarrow 2$ transitions
in the spectrum of surface electrons levitating above the surface of solid 
hydrogen in the
presence of static negative surface charge $\sigma $. Both the frequency shift 
and the line 
broadening were measured at fixed wavelengths of excitation light (79 $\mu $m 
and 84.3 $\mu $m) by
tuning the confining electric field $E$. Energy spectrum (4) must be used to 
convert these data 
into the units of frequency. The dashed line has a slope of $2\pi $.
}
\label{fig1}
\end{figure}

Similar negative frequency shift was observed also in the case of positive 
static surface charge,
although detailed measurements of the frequency shift and line broadening were 
not conducted in
this case. It was possible to create a two-dimensional electron layer on the 
surface of positively 
charged hydrogen crystal (in zero or even repulsive external electric field 
$E$). Both 
positive ions charging the surface and free electrons were deposited in turn on 
the surface of 
hydrogen crystal at P=5 Torr hydrogen vapor pressure from the gas discharge 
created in the 
experimental chamber. 
The life time of free electrons in such a system was longer than an hour which 
allowed us to
perform spectroscopic measurements shown in Fig.2. Photoresonance $1\rightarrow 
2$ transition has 
been detected in such a system at $\lambda =118.6 \mu m$ wavelength of 
excitation light
at $N\sim 10^{10} cm^{-2}$ density of positive ions deposited onto the hydrogen 
crystal 
surface. This
corresponds to the frequency of $1\rightarrow 2$ transition substantially below 
the frequencies
of $1\rightarrow 2$ transitions in zero confining field observed in surface 
electron layers in
the absence of static charge in all the temperature range studied (7-13.8 K). 
The fact that negative frequency
shift is observed both for positive and negative static surface charge is a 
strong argument
in favor of electric field induced liquid hydrogen supercooling phenomenon.

\begin{figure}[tbp]
\centerline{
\psfig{figure=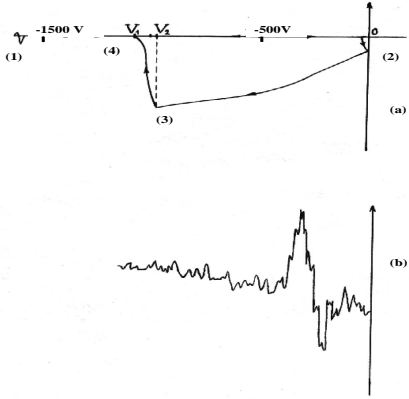,width=8.0cm,height=7.4cm,clip=}
}
\caption{ Photoresonance $1\rightarrow 2$ transition detected at $\lambda =118.6 
\mu m $ wavelength
of excitation light in the two-dimensional electron layer levitating above the 
surface of 
positively charged $\sim 0.5$ mm thick solid hydrogen crystal. The signal (b) 
proportional to the
derivative of the optical absorption was recorded while scanning the potential 
of the electrode 
located below the hydrogen crystal (note that the external field applied to the 
electrode is 
repulsive for the surface electrons). The upper curve (a) show the non resonant 
radio frequency 
absorption of the electron layer used to characterize the electron density [8]. 
Positive ions
charge the hydrogen surface at point (1) when gas discharge was created in the 
experimental 
chamber. When the potential of the electrode was reversed, the gas discharge 
appeared again at
point (2) leading to the appearance of free electrons on the hydrogen surface. 
The curve (b)
was measured on the way from (2) to (3). At point (3) free electrons started to 
leave the hydrogen
surface. They completely disappeared at point (4).
}
\label{fig2}
\end{figure}

Although the physical location of static charges with respect to the hydrogen 
surface is not as 
clear as in the case of free electrons (and, hence, the sample is not as well 
characterized) our 
data show that the charged solid hydrogen phase is easily accessible to 
experiments, and it is a
very interesting physical object from the point of view of superfluid molecular 
hydrogen 
observation. If $E_{\lambda }$ is a field necessary for supercooled liquid and 
solid phase
coexistence, the diameter of the area around a static charge where this 
condition is met
is determined by $d=(4e/E_{\lambda })^{1/2}$. Using the result of \cite{7} one 
may get $d\sim 15\AA $.
At static charge concentration of $N\sim 1/(4d^2)= 10^{13} cm^{-2}$ one may 
expect a substantial 
portion of the crystal surface to be in adequate condition for the liquid and 
solid phase 
coexistence at T=1 K. It must be also noted that the values of temperature and 
field 
necessary for the superfluid transition to occur are not quite known. In reality 
one may
see the onset of superfluid transition at sufficiently lower static charge 
densities. 
 
In conclusion, it was pointed out that the electric fields comparable to the 
field which
may stabilize supercooled liquid hydrogen at the temperature of superfluid 
transition
occur naturally in such existing physical systems as electrons levitating above 
the surface
of solid hydrogen and statically charged hydrogen crystals. The results of 
photoresonance 
measurements performed on the system of levitating electrons with and without 
static
surface charge strongly indicate the possibility of liquid hydrogen supercooling 
in
these systems. It would be extremely interesting to study the behavior of 
charged solid
hydrogen at temperatures around T=1 K with the goal of the observation of 
supercooled
superfluid hydrogen phase.

\end{document}